\documentclass[conference]{IEEEtran}

\usepackage{url}
\usepackage{graphicx}
\usepackage{balance}
\usepackage{graphics}
\usepackage{graphicx}
\usepackage{epsfig,amssymb}
\usepackage{epstopdf}
\usepackage{times}
\usepackage{mathrsfs}
\usepackage{array}
\usepackage{amsmath, latexsym, amsfonts, amssymb,bm}
\usepackage[mathscr]{eucal}
\usepackage{array, tabularx}
\usepackage[table]{xcolor}
\usepackage{multirow}
\usepackage{epsfig}
\usepackage{cite}
\usepackage{tikz}
\usepackage{mathtools}
\usepackage{psfragx}
\usepackage{xcolor}
\usepackage{bbm}
\usepackage{lineno}

\usepackage{algorithm}
\usepackage{algorithmic}

\usepackage{lipsum}

\usepackage{caption}
\usepackage{subcaption}
\usepackage{authblk}

\newcommand{\defeq}{\ensuremath{\triangleq}} 
\newcolumntype{S}{>{\centering\arraybackslash}p{5em}}

\newtheorem{proposition}{Proposition}

\DeclareMathOperator{\diag}{{diag}}

\IEEEoverridecommandlockouts

\begin{document}

\title{Intelligent Reflecting Surface: Practical Phase Shift Model and Beamforming Optimization}

\author[$\ast \dag $]{Samith Abeywickrama}
\author[$ \ast  $]{Rui Zhang}
\author[$ \dag $]{Chau Yuen}
\affil[$ \ast $]{National University of Singapore, Singapore}
\affil[$ \dag $]{Singapore University of Technology and Design,  Singapore}

\affil[$  $ ]{Email: \textit { samith@u.nus.edu, elezhang@nus.edu.sg, yuenchau@sutd.edu.sg} 

\thanks{A more comprehensive version of this work has been submitted to IEEE Transactions on Communications \cite{abeywickrama2019intelligent}}
}

\maketitle

\begin{abstract}
	Intelligent reflecting surface (IRS) that enables the control of the wireless propagation environment has been looked upon as a promising technology  for boosting  the spectrum and energy efficiency in future wireless communication systems.
	Prior works on IRS are mainly based on the ideal  phase shift model  assuming the  full signal  reflection by each of the elements regardless of its  phase shift, which, however, is practically difficult to realize. In contrast, we propose  in this paper a practical phase shift model that captures the phase-dependent amplitude variation  in the element-wise reflection coefficient. Applying  this new model to an IRS-aided wireless system, we formulate a problem to maximize its achievable rate by jointly optimizing the transmit beamforming and the IRS reflect beamforming. The formulated problem is non-convex and difficult to be optimally solved in 	general, for which we propose a low-complexity suboptimal solution based on the alternating optimization (AO) technique. Simulation results unveil a substantial performance gain  achieved by the joint beamforming optimization  based on the proposed phase shift model as compared to  the conventional ideal model.
\end{abstract}


\IEEEpeerreviewmaketitle

\section{Introduction} 

Intelligent reflecting surface (IRS) assisted wireless communication has recently emerged as a promising solution to enhance the spectrum and energy efficiency for future wireless systems. Specifically, an IRS is able to establish  favourable channel responses by controlling the wireless propagation environment  through its reconfigurable passive reflecting elements (see e.g. \cite{wu2018intelligent,qq_magazine,chongwang,8796365} and the references therein). However, the  existing works on IRS mostly  assume an ideal phase shift model with full reflection, i.e., unity amplitude at each reflection element regardless of the phase shift, which, however, is practically difficult to realize due to the hardware limitation.

The amplitude response of a typical passive reflecting element is non-uniform
with respect to its phase shift. In particular, the amplitude  exhibits its minimum value at the zero phase shift, but monotonically increases and asymptotically approaches  unity amplitude at the phase shift of $\pi$ or $-\pi$. This is due to the fact that when the phase shift approaches zero, the image currents, i.e., the currents of a virtual source that accounts for the reflection, 
are in-phase with the reflecting element currents, and thus the electric field and the current flow in the element are  enhanced. As a result, the dielectric loss, metallic loss, and ohmic loss  increase dramatically, leading to  substantial energy loss and hence low reflection amplitude \cite{4619755}. Furthermore, these losses mainly come from the semiconductor devices, metals, and dielectric substrates used in the IRS, and thus are  not avoidable in practice. In fact,  this is a long standing problem for reflection-based metasurfaces \cite{zhu2013active}.  
In \cite{irs_amp}, amplifiers are integrated into the reflecting elements to compensate the energy loss, which is not suitable for passive IRS and also practically costly.

In \cite{wu2018intelligent} and \cite{chongwang}, by assuming the ideal phase shift model,  IRS reflection is designed to have the maximum phase alignment between the IRS-reflected and non-IRS-reflected signals at the designated receivers.
In contrast, when the amplitude depends on the phase shift  at each reflecting element, such an optimal reflection design is not feasible  as each phase shift needs to be properly chosen to have a better balance between the amplitude and phase alignment. Therefore, if the IRS reflection is designed for a practical system based on the ideal phase shift model, it inevitably causes certain performance degradation. To the best of the authors' knowledge, the practical  phase shift model and  corresponding beamforming optimization algorithm design for IRS-aided wireless systems has not been reported in the literature yet.

This thus motivates this paper, where  we first propose a practical  phase shift  model and verify its accuracy with the experimental results reported in literature. Next, based on this model and considering  an IRS-aided point-to-point  communication  system, we formulate a new problem to maximize its achievable rate by jointly optimizing the transmit beamforming and the IRS reflect beamforming.  As this problem is non-convex, we propose a low-complexity algorithm to solve it
sub-optimally by leveraging the alternating optimization (AO)
technique.   Simulation results are also presented to demonstrate  the performance gain by the joint beamforming optimization based on the proposed practical phase shift model over the conventional ideal model.

\textit{Notations:} In this paper, scalars are denoted by italic letters, vectors and matrices are denoted by bold-face lower-case and upper-case letters, respectively. For a complex-valued vector $ \mathbf v $, $ \| \mathbf v \| $, $ \mathbf v^H $,  and $\diag(\mathbf v)$ denote its $\ell_2$-norm, conjugate transpose, and a diagonal matrix with each diagonal element being the corresponding element in $ \mathbf v $, respectively. Scalar $v_i$ denotes the $i$-th element of vector $\mathbf v$. For a square matrix $ \mathbf{A} $, $\mathbf{A}_{n,k}$ denotes its entry in the $n$-th row and $k$-th column.  $ \mathbb{C}^{x \times y} $  denotes the space of $ x \times y $ complex-valued matrices. $ j $ denotes the imaginary unit, i.e., $ j^2 = -1 $.  For a complex-valued scalar $  v $, $ |  v | $, $\arg(v)$, and $ \bar v $ denote its absolute value, phase, and complex conjugate, respectively. $\mathbb{E}(\cdot)$ denotes the statistical expectation.

\section{System Model}

\begin{figure}[t!]
	\centering \vspace{1mm}
	\includegraphics[width=0.7\linewidth]{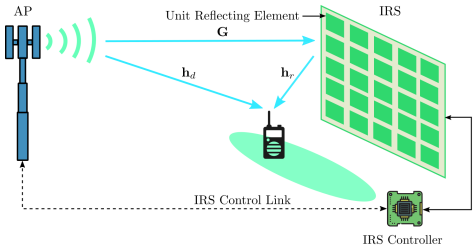}
	\caption{{An IRS-aided wireless system.}}  \vspace{-2mm}
	\label{system_model} 
\end{figure}

We consider a multiple-input single-output (MISO) wireless system where an IRS composed of $N$ reflecting elements is deployed to assist in the communication from an access point (AP) with $M$ antennas to a single-antenna user,  as illustrated in Fig. \ref{system_model}. The IRS reflecting elements are  programmable via an IRS controller. Furthermore, IRS controller communicates with the AP via a separate wireless link for the AP to control the IRS reflection. 
It is  assumed that the signals that are reflected by the IRS more than once  have negligible power due to substantial path loss and  thus are ignored. 
In addition, we consider a quasi-static flat-fading model, where it is assumed that all the wireless channels remain constant over each transmission block. The channels are assumed to be known at the AP by applying, e.g., the channel estimation technique proposed in \cite{IRS_channel_estimation}.  

Let $\mathbf h_d \in \mathbb C^{M\times1}$, $\mathbf h_r \in \mathbb C^{N\times1}$, and $\mathbf G \in \mathbb C^{N\times M}$ denote the  baseband equivalent channels from the AP to user, from the IRS to user, and from the AP to IRS, respectively. Without loss of generality, let  $\mathbf v  \in \mathbb C^{N\times1}$   denote the reflection coefficient vector of the IRS, where $|v_n| \in [0,  1]$ and $ \arg(v_n) \in [-\pi,  \pi)$  are the amplitude and the phase shift on the combined incident signal, respectively, for $n \in \{1,\dots,N\} $ \cite{qq_magazine}. Note that for the ideal phase shift model considered in \cite{wu2018intelligent,qq_magazine,chongwang}, $|v_n|=1,\forall n$, regardless of the phase shift, $\arg(v_n)$. The  transmit signal at the AP is given by $ \mathbf{x} = \mathbf{w} s $, where $\mathbf w \in \mathbb{C}^{M \times 1}$ denotes the beamforming vector and $ s $ denotes the transmit symbol, which is independent of $\mathbf{w}$, and has zero-mean and unit variance ({i.e.,} $ \mathbb{E}( |s|^{2}) =1 $). We have dropped the time index for notational simplicity.  The received baseband signal at the user is thus given by 
\begin{align} \label{received_signal}
y &=  (\mathbf v^H  \mathbf \Phi + \mathbf h_d^H)\mathbf{x} + z,
\end{align} 
where $\mathbf \Phi=   \diag(\mathbf h_r^H) \mathbf G$ and $z$ denotes the additive white Gaussian noise (AWGN) at the receiver with zero mean and variance $\sigma^2$.

In this paper, we aim to maximize the achievable rate or spectrum efficiency (SE) in bits per second per Hertz (bps/Hz) by jointly optimizing the AP beamforming vector $\mathbf{w}$ and the IRS reflection vector $\mathbf v$. Accordingly, the achievable rate/SE  is given by\footnote{Note that the considered  system model can be also applied to  wireless power transfer (WPT) \cite{qq_magazine} as the harvested radio-frequency (RF)  energy at the receiver is generally modeled as an  increasing function of the received signal power   \cite{7867826}, i.e., the term  $ |(\mathbf v^H  \mathbf \Phi + \mathbf h_d^H)\mathbf{w}|^2$ given in \eqref{r_se}.}
\begin{align} \label{r_se} 
R_{\text{SE}} &=  \log_2 \left( 1+ \frac{|(\mathbf v^H  \mathbf \Phi + \mathbf h_d^H)\mathbf{w}|^2}{\sigma^2}\right). 
\end{align} 

\section{Practical  Phase Shift Model}

\begin{figure}[t!]
	\centering \vspace{1mm}
	\includegraphics[trim = 0mm 0mm 0mm 0mm, clip,width=0.8\linewidth]{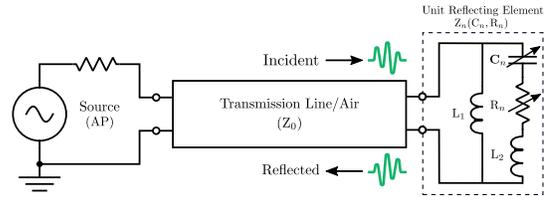}
	\caption{{Transmission line model of a unit reflecting element.}}  
	\label{transmission_model} 
\end{figure}


\subsection{Equivalent Circuit Model} \label{circuit_sec}





An IRS is typically constructed as a printed circuit board (PCB), where the reflecting elements are equally spaced  in a two-dimensional plane. 
A unit reflecting element is composed of a metal patch on the top layer of the PCB dielectric substrate and a full metal sheet on the bottom layer \cite{qq_magazine}. 
Moreover, a semiconductor device\footnote{In practice, a positive-intrinsic-negative (PIN) diode, a variable capacitance (varactor) diode, or a metal-oxide-semiconductor field-effect transistor (MOSFET) can be used as the semiconductor device mentioned here  \cite{tang2018wireless,zhu2013active,PhysRevApplied}. }, which can vary the impedance of the reflecting element by controlling its biasing voltage, is embedded into the top layer metal patch  so that the element response  can be dynamically tuned in real time without changing the geometrical parameters \cite{PhysRevApplied}. In other words, when the geometrical parameters are fixed, the semiconductor device  controls the phase shift and amplitude (absorption level).



As the physical length of a unit reflecting element is usually smaller than  the wavelength of the desired incident  signal, its response can be accurately
described by an equivalent lumped circuit model regardless of the particular
geometry of the element \cite{koziel2013surrogate}.
As such, the metallic parts in the reflecting element can be modeled as inductors as the high-frequency current flowing on it produces a quasi-static magnetic field. In  Fig. \ref{transmission_model}, the equivalent model for the $n$-th reflecting element is illustrated  as a parallel resonant circuit and its impedance is given by
\begin{align} \label{z_n}
Z_n(C_n,R_n) &=  \frac{j \omega L_1 (j\omega L_2+\frac{1}{j \omega C_n}+R_n) }{j \omega L_1 + (j\omega L_2+\frac{1}{j \omega C_n}+R_n)},
\end{align}
where $L_1$, $L_2$, $C_n$, $R_n$, and $\omega$ denote the bottom layer inductance, top layer inductance, effective capacitance, effective resistance, and angular frequency of the incident signal, respectively. 
Note that $R_n$ determines the amount of power dissipation due to the losses in the semiconductor devices, metals, and dielectrics, which cannot be zero in practice, and $C_n$ specifies  the charge accumulation related to  the element geometry and semiconductor device.
As the transmission line diagram in Fig. \ref{transmission_model} depicts, the reflection coefficient, i.e., $v_n$ in \eqref{received_signal}, is the parameter that describes the fraction of the reflected electromagnetic wave  due to the impedance discontinuity between the free space impedance  $Z_0$ and element impedance $Z_n(C_n,R_n)$  \cite{microwave_book}, which  is given by  
\begin{align} \label{v_n}
v_n &=  \frac{Z_n(C_n,R_n) - Z_0}{Z_n(C_n,R_n) + Z_0}.
\end{align}
Since $v_n$ is a function of $C_n$ and $R_n$, the reflected electromagnetic waves can be manipulated  in a controllable and programmable manner by varying $C_n$'s and $R_n$'s.

\begin{figure} 
	\centering
	\begin{subfigure}[b]{0.32\textwidth} 
		\centering
		\includegraphics[trim = 0mm 0mm 0mm 0mm, clip,width=\columnwidth]{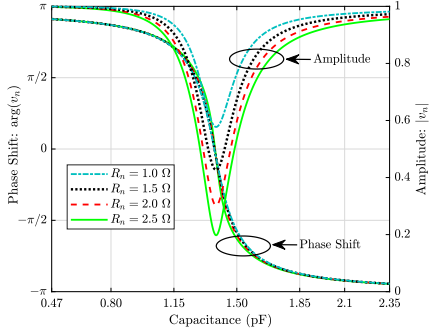}
		\caption{Phase shift and amplitude  versus  $C_n$ and $R_n$.}
		\label{fig_3_a} 
	\end{subfigure}
	\hspace{-0.1cm}
	\begin{subfigure}[b]{0.32\textwidth} 
		\centering
		\includegraphics[trim = 0mm 0mm 0mm 0mm, clip,width=\columnwidth]{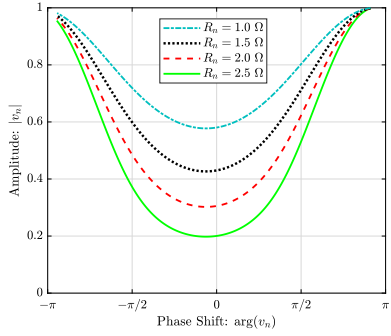}
		\caption{Amplitude versus  phase shift. }
		\label{fig_3_b} 
	\end{subfigure}
	\caption{Reflection coefficient of a unit reflecting element.} \label{reflection} 
\end{figure} 

 \begin{figure} 
 	\centering
 	\begin{subfigure}[b]{0.32\textwidth} 
 		\centering
 		\includegraphics[trim = 0mm 0mm 0mm 0mm, clip,width=\columnwidth]{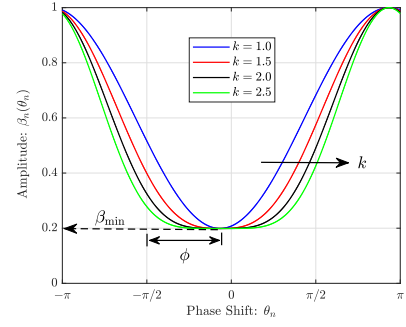}
 		\caption{The phase shift model with different parameters.}
 		\label{fig_4_a} 
 	\end{subfigure}
 	\hspace{-0.1cm}
 	\begin{subfigure}[b]{0.32\textwidth} 
 		\centering
 		\includegraphics[trim = 0mm 0mm 0mm 0mm, clip,width=\columnwidth]{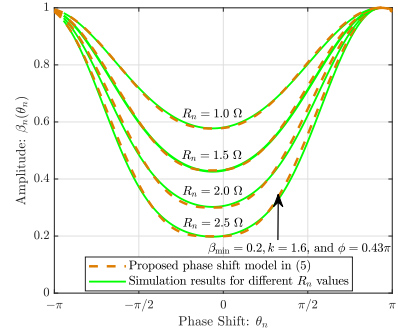}
 		\caption{Simulation results for the proposed  phase shift model. }
 		\label{fig_4_b} 
 	\end{subfigure}
 	\caption{The proposed phase shift model.} \label{model} 
 \end{figure}

To demonstrate this, Fig. \ref{reflection} illustrates the behaviour of the amplitude  and the phase shift, i.e., $ |v_n|$ and $\arg(v_n)$, respectively, for different values of $C_n$ and $R_n$.  Note that to align with the experimental results in \cite{zhu2013active}, $C_n$ is varied from $0.47$ pF to $2.35$ pF when $L_1=2.5$ nH, $L_2=0.7$ nH, $Z_0=377 \ \ \Omega$, and $\omega=2\pi \times 2.4 \times 10^9$. It is observed that a reflecting element is capable of achieving almost $2\pi$ full phase tuning, while the phase shift and amplitude both  vary with  $C_n$ and $R_n$ in general. It is also observed that the minimum amplitude occurs  near zero  phase shift and  approaches unity (the maximum)  at the phase shift of $\pi$ or $-\pi$, which is explained as follows. When the phase shift is around $\pi$ or $-\pi$, the reflective currents (also termed as image currents) are out-of-phase with the element currents, and thus the electric field  and the current flow in the element are both  diminished, thus resulting in minimum  energy loss and highest  reflection amplitude. In contrast, when the phase shift is around  zero, the reflective currents are in-phase with the element currents, and thus the electric field  and the current flow in the element are both  enhanced. As a result, the dielectric loss, metallic loss, and ohmic loss increase dramatically, leading to  substantial energy dissipation and thus lowest reflection amplitude.
Furthermore, it is worth noting that the numerical results illustrated in Fig. \ref{reflection} are in accordance with  the experimental results reported in literature (see \cite{4619755} and Fig. 5 (b) in \cite{zhu2013active}), indicating that the circuit model given by \eqref{z_n} and \eqref{v_n}  accurately captures the physics of a reflecting element in practice.

It is also  worth noting that to obtain an ideal phase shift control, where $|v_n| = 1,\forall \arg(v_n) \in [-\pi,  \pi)$, each reflecting element should exhibit zero energy dissipation. However, for practical hardware, energy dissipation is unavoidable\footnote{In \cite{zhu2013active}, $R_n=2.5$ $\Omega$ in each reflecting element due to the diode junction resistance, while in  \cite{4619755}, although the reflecting element does not contain any semiconductor device, its amplitude response follows a similar shape to  Fig. \ref{reflection} due to the metallic loss and dielectric loss.} and the typical behaviour of the reflection amplitude is similar to Fig. \ref{reflection}.   Therefore, incorporating a practical phase shift model to design  beamforming  algorithms  is essential to optimize  the performance of  IRS-aided wireless systems.

\subsection{Proposed Phase Shift Model}

In order to characterize  the fundamental  relationship between the reflection  amplitude and  phase shift for designing IRS-aided wireless systems, we propose in this subsection an analytical model for the phase shift which is generally  applicable to a variety of semiconductor devices used for implementing the IRS. Let $v_n = \beta_n(\theta_n) e^{j \theta_n} $ with $\theta_n \in [-\pi,  \pi)$ and $\beta_n(\theta_n) \in [0,  1]$ respectively  denote  the phase shift and the corresponding amplitude. Specifically, $\beta_n(\theta_n)$ can be expressed  as \vspace{-2mm}
\begin{align} \label{our_model}
\beta_n(\theta_n) &=  (1-\beta_{\text {min}}) \left( \frac{\sin(\theta_n - \phi) +1}{2} \right)^k + \beta_{\text {min}},
\end{align}
where $\beta_{\text {min}}\geq 0$, $\phi \geq 0$, and $k\geq0$ are the constants related to the specific circuit implementation. As depicted in Fig. \ref{model} (a), $\beta_{\text {min}}$ is the minimum amplitude, $\phi$ is the  horizontal distance between $-\pi/2$ and $\beta_{\text {min}}$,  and $k$ controls the steepness of the function curve. Note that for $k=0$, \eqref{our_model} is equivalent to the ideal phase shift model, i.e., $\beta_n(\theta_n)=1,\forall n$. In practice, IRS  circuits are fixed once they are fabricated and these parameters can be easily found by a standard curve fitting tool.

Fig. \ref{model} (b) illustrates that the proposed phase shift model  closely matches the simulation results presented in Section \ref{circuit_sec} for a practical reflecting element. In the sequel, we adopt the model in \eqref{our_model}  for beamforming design in IRS-aided wireless communication. Moreover, we assume that the  circuits of the reflecting elements are all identical, and thus the same model parameters, i.e., $\beta_{\text {min}}$, $\phi$, and $k$,  apply to  each of the elements.

\section{Beamforming Optimization}

\subsection{Problem Formulation}
We aim  to jointly optimize  $\mathbf{w}$ and $\mathbf{v}$ such that the achievable rate,  $R_{\text{SE}}$ given in \eqref{r_se}, is maximized. The  problem is formulated as 
\begin{align} 
\mathrm{(P0)}:  
\mathop{\mathtt{max}}_{\mathbf{w},\mathbf{v}, \{\theta_n\}}~& |(\mathbf v^H  \mathbf \Phi + \mathbf h_d^H)\mathbf{w}|^2  \label{eq:P0_Obj} \\
\mathtt{s.t.}~&  \|\mathbf{w}\|^2_2 \leq P_T, \label{eq:P0_C1} \\
~& v_n = \beta_n(\theta_n) e^{j \theta_n}, \forall n = 1,\dots,N, \label{eq:P0_C2} \\
~& -\pi \leq \theta_n \leq \pi, \forall n = 1,\dots,N, \label{eq:P0_C3}
\end{align}
where  $P_T$ denotes the maximum transmit power constraint at the AP. For any given phase shift $\mathbf v$, it is known that the maximum-ratio transmission (MRT) is the optimal transmit beamforming solution to (P1), i.e., $\mathbf{w}^*=\sqrt{P_T}\frac{(\mathbf v^H  \mathbf \Phi + \mathbf h_d^H)^H}{\|(\mathbf v^H  \mathbf \Phi + \mathbf h_d^H)\|}$. By substituting $\mathbf{w}^*$ to (P0), the problem for optimizing the IRS reflection is reformulated as 
\begin{align} 
\mathrm{(P1)}:  
\mathop{\mathtt{max}}_{\mathbf{v},\{\theta_n\}}~& \|(\mathbf v^H  \mathbf \Phi + \mathbf h_d^H)\|^2  \label{eq:P1_Obj} \\
\mathtt{s.t.}~&  v_n = \beta_n(\theta_n) e^{j \theta_n}, \forall n = 1,\dots,N, \label{eq:P1_C1} \\
~& -\pi \leq \theta_n \leq \pi, \forall n = 1,\dots,N. \label{eq:P1_C2}
\end{align}
Although simplified, problem (P1) is non-convex and difficult to be optimally solved
in general. In the next subsection, we solve (P1) by applying the AO technique.

\subsection{Proposed AO Algorithm}

We propose an AO algorithm to find an approximate solution to (P1), by iteratively optimizing the phase shift of one of the $N$ reflecting elements with those of the others  being fixed at each time, and repeatedly doing this procedure for all $N$ elements until the objective value in \eqref{eq:P1_Obj} converges. The convergence  is guaranteed as the optimal value of (P1) is upper-bounded by a finite value. To this end, the problem for optimizing the reflection of the $n$-th element is formulated as 
\begin{align} 
\mathrm{(P2)}:  
\mathop{\mathtt{max}}_{\theta_n}~& \beta_n^2(\theta_n) \mathbf{\Psi}_{n,n} + \beta_n(\theta_n)|\varphi_n| \cos( \arg(\varphi_n)-\theta_n )  \label{eq:P1_AO_Obj}  \\
\mathtt{s.t.}~&  -\pi \leq \theta_n \leq \pi, 
\end{align}
where $\mathbf{\Psi}=\diag(\mathbf h_r^H)\mathbf G\mathbf G^H\diag(\mathbf h_r)$, $\mathbf {\hat h}_d =\diag(\mathbf h_r^H)\mathbf G \mathbf h_d$, and $\varphi_n = \left( \sum_{m\neq n}^{N} \mathbf{\Psi}_{n,m}  v_m \right) +  2\hat{h}_{d,n} $. Note that \eqref{eq:P1_AO_Obj} is obtained by taking the terms associated with  $\beta_n(\theta_n)$ and $\theta_n$ in the expansion of \eqref{eq:P1_Obj}, while   the derivation is omitted due to the space limitation. The problem (P2) is a single-variable non-convex optimization problem, for which we propose a closed-form approximate solution that can be efficiently computed in the next subsection.

\subsection{An Approximate Solution to (P2)}

The key to approximately solve  (P2) in closed-form lies in re-expressing \eqref{eq:P1_AO_Obj} in a more tractable model.  However, an approximate model of a general nonlinear function can only fit the original function locally, which we refer to as the trust region. In our problem, the trust region should essentially be the one that encloses the optimal solution of (P2), denoted by $\theta_n^*$.

Define $f(\theta_n) \defeq \beta_n^2(\theta_n) \mathbf{\Psi}_{n,n} + \beta_n(\theta_n)|\varphi_n| \cos( \arg(\varphi_n)-\theta_n )$. It is not difficult to observe that for the ideal phase shift model considered in \cite{wu2018intelligent,qq_magazine,chongwang},  $\beta_n(\theta_n)$ and $\theta_n$  can be  designed to maximize $f(\theta_n) $ (or \eqref{eq:P1_AO_Obj})  by setting $\beta_n^*(\theta_n) =1$ and $\theta_n^*=\arg(\varphi_n), \forall n$. However, such an optimal reflection design is not feasible  for a  practical IRS due to  the dependency of $\beta_n(\theta_n)$ on $\theta_n$ as depicted in Fig. \ref{reflection} (b). For instance, if $\arg(\varphi_n) = 0$, $\theta_n^* = 0$ may not be a favourable phase design as it yields  the lowest  reflection amplitude. In this case, $\theta_n^*$ needs to be properly chosen to have a better balance between $\beta_n(\theta_n) $ and  $\arg(\varphi_n)$. In particular, since the minimum $\beta_n(\theta_n)$ occurs  near zero  phase shift and  approaches the maximum  at  $\pi$ and $-\pi$, $\theta_n^*$ should slightly deviate from $\arg(\varphi_n)$ towards  $\pi$ (or $-\pi$) when $\arg(\varphi_n)$ is non-negative (negative). The trust region that encloses $\theta_n^*$ is thus given by 
\begin{align} 
\theta_n^* \in [\arg(\varphi_n),(-1)^\lambda\pi], \label{t_r}
\end{align}
with $\lambda =0$ when $\arg(\varphi_n) \geq 0$  and $\lambda =1$ otherwise. 

Motivated by the above result, a high-quality approximate solution to problem (P2) can be obtained numerically via a one-dimensional (1D) search over $[\arg(\varphi_n),(-1)^\lambda\pi]$, which, however, is still computationally inefficient. Alternatively, a closed-form approximate solution  can be obtained by fitting a quadratic model through three points over the trust region (which are obtained via equally sampling the trust  region), i.e., $\theta_A=\arg(\varphi_n)$, $\theta_B=\frac{\arg(\varphi_n)+(-1)^\lambda\pi}{2}$, and $\theta_C=(-1)^\lambda\pi$, as given in the following  proposition. 
\begin{proposition} \label{sub_allocation_pro}
	Let	$f_1 = f(\theta_A)$, $f_2 = f(\theta_B)$, and	$f_3 = f(\theta_C)$. The approximate solution to (P2) obtained by fitting a quadratic curve through the points $(\theta_A,f_1)$, $(\theta_B,f_2)$, and $(\theta_C,f_3)$ is given by
	\begin{align} \label{our_method}
	\hat\theta_n^* = \frac{(-1)^\lambda\pi(3f_1-4f_2+f_3) +\arg(\varphi_n)(f_1-4f_2+3f_3)  }{4(f_1-2f_2+f_3)}.
	\end{align}
\end{proposition} 
\begin{IEEEproof}
	See Appendix \ref{single_ER}.
\end{IEEEproof}
It is worth noting that Proposition 1 essentially corresponds to a single iteration of successive quadratic estimation with trust region refinement \cite{powell}. 
The overall iterative algorithm to solve (P1) is given in Algorithm 1.

\begin{algorithm} [h!] 
	\caption{Alternating Optimization (AO) for Solving (P1)}\label{Tabel_2} 
	\begin{algorithmic}[1]
		\STATE \textbf{Initialize:} {  $\{\theta_n\}_{n=1}^N$ }
		\REPEAT
		\FOR{$n=1$ \textbf{to} $N$}
		\STATE Find $\theta_n^*$ as the solution to (P2).
		\ENDFOR
		\STATE Obtain $v_n = \beta_n(\theta_n^*) e^{j \theta_n^*},\forall n$.
		\UNTIL{{the objective value of (P1) with the obtained  $\mathbf{v}$ reaches convergence;}}
	\end{algorithmic}
\end{algorithm} 

\section{Simulation Results} \label{simulation}

\begin{figure}[t!] 
	\centering
	\includegraphics[trim = 0mm 0mm 0mm 0mm, clip,width=0.75\linewidth]{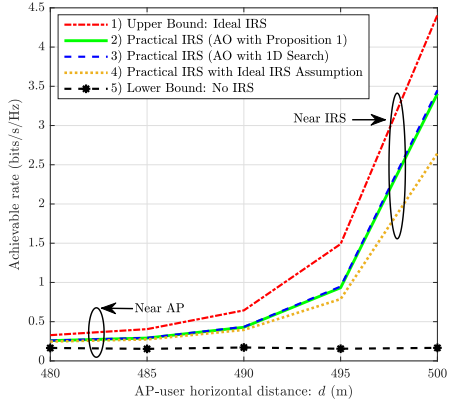}
	\caption{Achievable rate versus the AP-user horizontal distance  when $N=40$.}  
	\label{variable_d} 
\end{figure} 

We consider a MISO downlink wireless system consisting of an AP with $M=2$ antennas and a single-antenna user. It is assumed that an IRS composed of $N$ reflecting elements is deployed in the vicinity of the  user while the AP and IRS are assumed to be located $500$ meters (m) apart. Rayleigh fading is assumed for all the channels involved, and the signal attenuation at a reference distance of $1$ m is set as $40$ dB. The path loss exponents are set to $2.2$, $2.8$, and $3.8$ for the channels between AP-IRS, IRS-user, and AP-user, respectively, according to\cite{wu2018intelligent}. The total transmit power at the AP is $P_T=36$ dBm and $\sigma^2=-94$ dBm. 

The user is assumed to lie  on a horizontal line that is in parallel to  that connecting the  AP and IRS, with the vertical distance between these two lines equal to $2$ m. By varying  the horizontal distance between the AP and user, denoted by $d$, in Fig. \ref{variable_d}, the achievable rate averaged  over $1000$ channel realizations is shown  for the following schemes: 
\begin{enumerate} 
	\item Upper bound: solving the following problem, which assumes the ideal phase shift model, and its solution is given in \cite{wu2018intelligent}.
	\begin{align} 
	\mathrm{(P3)}:  
	\mathop{\mathtt{max}}_{\mathbf{\tilde v}}~& \|(\mathbf {\tilde v}^H  \mathbf \Phi + \mathbf h_d^H)\|^2  \label{eq:P4_Obj} \\
	\mathtt{s.t.}~&  |\tilde v_n|^2 = 1, \forall n = 1,\dots,N. 
	\end{align} 
	\item Beamforming optimization by  the  AO algorithm under the proposed practical phase shift model with  $\beta_{\text{min}}=0.2$, $k=1.6$, and $\phi=0.43\pi$, while the problem (P2) is solved using Proposition 1.
	\item Beamforming optimization by  the  AO algorithm under the proposed practical phase shift model with  $\beta_{\text{min}}=0.2$, $k=1.6$, and $\phi=0.43\pi$, while the problem (P2) is solved using 1D search.
	\item Beamforming optimization  assuming the ideal  phase shift model \cite{wu2018intelligent},  while  the practical  phase shift model is used for computing the achievable rate. 
	\item Lower bound: the system without using an IRS by setting
	$ \mathbf w^* =\sqrt{P_T}\frac{\mathbf h_d}{\|\mathbf h_d\|}$. 
\end{enumerate}
Note that the initial phase shift values of the proposed AO algorithm, i.e., $\{\theta_n\}_{n=1}^N$, are randomly selected from $\{\pi,-\pi\}$ such that each reflecting element has the maximum reflection amplitude. 

\begin{figure}[t!] 
	\centering
	\includegraphics[trim = 0mm 0mm 0mm 0mm, clip,width=0.75\linewidth]{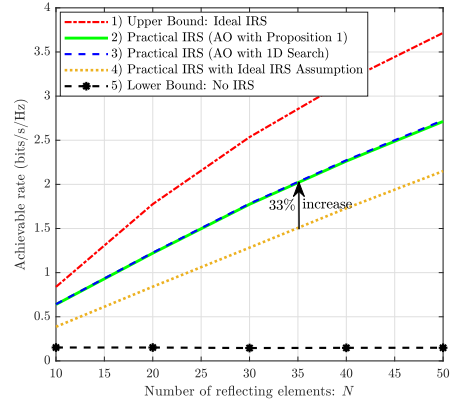}
	\caption{Achievable rate versus  number of reflecting elements at $d=498$ m.}  
	\label{variable_n} 
\end{figure} 

It is observed from Fig. \ref{variable_d} that 2) performs very close to 3).  Proposition 1 thus provides a practically appealing solution to (P2) considering its performance and low complexity. It is also observed that when the user moves closer to the IRS, the performance gap between 2) and 4) increases. This is due to the fact that the user benefits from the stronger reflecting channel via IRS ($\mathbf h_r$), and therefore accurate reflection design at the IRS becomes more  crucial. In contrast, when the user moves toward  the AP, the performance gap between 2) and 4) decreases as the  AP-user direct channel ($\mathbf h_d$) becomes dominant and the effect of IRS reflection becomes less significant. Moreover, by fixing the user at $d=498$ m and varying the number of reflecting elements, $N$, in Fig. \ref{variable_n}, we plot the average achievable rate. It is also observed that the performance gap between 2) and 4) increases with $N$ as the IRS reflecting channel becomes stronger. 

\begin{figure}[t!] \vspace{-3mm}
	\centering
	\includegraphics[trim = 0mm 0mm 0mm 0mm, clip,width=0.75\linewidth]{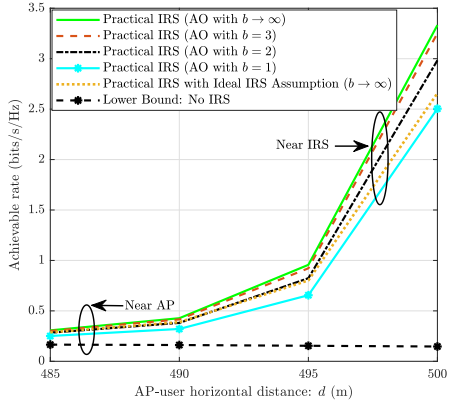}
	\caption{Achievable rate versus the AP-user horizontal distance  when $N=40$ in the case of discrete phase shift.}  
	\label{variable_d_d} \vspace{-4mm}
\end{figure} 

Next, we consider that the phase shift at each element of the IRS can only take a finite number of discrete values, which are equally spaced in $[-\pi,\pi)$ \cite{IRS_discrete}. Denote by $b$ the number of bits used to represent each of the levels. Then the set of phase shifts at each element is given by $\mathcal F = \{0,\Delta \theta,\dots,\Delta \theta(K-1)\}$ where $\Delta \theta = 2\pi/K$ and $K=2^b$.  When the user moves closer to the IRS, in Fig. \ref{variable_d_d}, we compare the average achievable rate for different values of $b$ with the practical and ideal phase shift model. Note that for the finite values of $b$, problem (P2) is solved by performing a 1D search over $\mathcal F$. As expected, the performance increases with $b$. Moreover, it is observed  that beamforming optimization under the practical phase shift model with $b=2$ performs even better than that of ideal phase shift model with $b \rightarrow \infty$.

\section{Conclusion} 

In this paper, we proposed  a practical IRS phase shift model. Based on   this new model and considering an IRS-aided MISO system, we formulated and solved  a joint transmit and reflect beamforming optimization problem to maximize the achievable rate, by applying the AO technique.  Our simulation results validated our proposed analytical model and showed  that beamforming optimization  based on the conventional ideal phase shift model, which has been  widely used in the literature, may lead to significant performance loss as compared to the proposed practical model. In future work, it is worth investigating such performance difference in more general IRS-aided wireless communication setups, such as multi-user systems \cite{wu2018intelligent,chongwang}, OFDM-based system \cite{irs_ofdm}, physical layer security system \cite{8723525}, simultaneous wireless information and power transfer (SWIPT) systems \cite{IRS_SWIPT_2,IRS_SWIPT_1}, and so on.

\appendices

\section{Beamforming Optimization} \label{single_ER}

\subsection{Proof of Proposition \ref{sub_allocation_pro}}  
Given three  points $\theta_A=\arg(\varphi_n)$, $\theta_B=\frac{\arg(\varphi_n)+(-1)^\lambda\pi}{2}$,  $\theta_C=(-1)^\lambda\pi$ and their corresponding function values $f_1$, $f_2$, $f_3$, we seek to determine three constants $a_0$, $a_1$, and $a_2$ such that the following  quadratic function is constructed,  
\begin{align} \label{q_f}
g(\theta_n) = a_0 + a_1(\theta_n-\theta_A) + a_2(\theta_n-\theta_A)(\theta_n-\theta_B).
\end{align}
When $\theta_n=\theta_A$, $\theta_n=\theta_B$, and $\theta_n=\theta_C$, the constants $a_0$, $a_1$, and $a_2$ can be respectively obtained. Substituting them into the stationary point of $g(\theta_n)$, i.e., $\hat \theta_n^\star = \frac{\theta_A+\theta_B}{2} - \frac{a_1}{2a_2}$, allows us to obtain \eqref{our_method}. The proof  is thus completed. \vspace{3mm}

\bibliographystyle{IEEEtran}  
\footnotesize{\bibliography{bibfile}}
\end{document}